%% file: EIHLMR_QLM_arxiv.tex
\documentclass{article}
\usepackage{amsmath}
\usepackage{amsfonts}
\usepackage{amssymb}
\usepackage{fullpage}
\usepackage{ctable}
\usepackage{pictex}

\newcommand{\refeq}[1]{(\ref{#1})}

\newcommand{\cA}{{\mathcal{A}}}

\renewcommand{\geq}{\geqslant}

\newcommand{\cL}{{\cal L}} 

\newcommand{\cS}{{\cal S}}
\reversemarginpar

\numberwithin{equation}{section}

\newcommand{\beq}{\begin{equation}}
\newcommand{\eeq}{\end{equation}}
\newcommand{\p}{\partial}
\newcommand{\el}{{\mbox{\tiny{el}}}}
\newcommand{\ph}{{\mbox{\tiny{ph}}}}
\newcommand{\bx}{\mathbf{x}}

\begin{document}

%

\title{The Einstein-Infeld-Hoffmann legacy in mathematical relativity\\
II.  Quantum laws of motion\footnote{Rapport by the first author given 
                        at the 15th Marcel Grossmann Meeting on General Relativity, 
                        Univ. Roma 1 (La Sapienza), Rome, Italy, July 3, 2018. Updated and revised: May 2019.}}
\author{{\bf A. Shadi Tahvildar-Zadeh$^\dagger$ and Michael K.~H. Kiessling}\\
Dept. of Mathematics, Rutgers--New Brunswick\\
E-mail: $^\dagger$shadit@math.rutgers.edu, miki@math.rutgers.edu}
\maketitle
\begin{abstract}
We report on recent developments towards a relativistic quantum mechanical theory of motion for a fixed, finite number of electrons, photons, and their anti-particles, as well as its  possible generalizations to other particles and interactions.
\end{abstract}

\section{Introduction}

As reported in Part I, the goal of the classical part of our program is to formulate and analyze a generally covariant joint initial value problem for the motion of massive charged particles, together with the electromagnetic and gravitational fields that they generate.  We believe that proving local well-posedness, i.e.~existence and uniqueness of a solution to this initial value problem, and  continuous dependence of the solution on initial data, is a first necessary step towards the formulation of a deeper quantum theory of motion, for which the above could serve as the classical limit.  

In the second part of our report,  we expand on what such a relativistic quantum theory could look like, and present the results of our preliminary efforts towards achieving our ultimate goal, which is twofold:
(1) to provide an accurate account of empirical electromagnetism in terms of 
a special-relativistic $N$-body quantum theory of electrons, photons, and their anti-particles; and (2) to extend that effort to general relativity by taking gravitational effects into account. 

At the same time, inspired by Einstein's \cite{Ein1909a,Ein1909b} and Bell's  \cite{BellBOOKsec} writings, we want to find out whether it is possible to formulate such a theory as a generalization of the non-relativistic theory of de Broglie \cite{deBroglieSOLVAY} and Bohm \cite{Bohm52}
in which the quantum-mechanical wave function $\psi$ guides the actual motion of all the particles involved.

Our approach to a relativistic quantum theory is based on a number of working hypotheses. Before listing them, we summarize the assumptions that formed the basis of our approach to the {\em classical} theory of motion of charged particles, as was outlined in Part I of this report:
\begin{itemize}
\item[1.] All forms of matter are of a discrete (particulate) nature.
\item[2.] Particles, whether feature-less point-particles or structured ones such as ring-particles, have well-defined world-lines or world-tubes that form the {\em singular boundary} of spacetime, in the sense that the spacetime metric and/or classical fields permeating the spacetime develop a singularity as a point on this boundary is approached. 
Moreover, these singularities must be weak enough to give rise to locally integrable field energy-momentum densities, but not so weak as
to be confined to characteristic hypersurfaces (which are completely determined by the field equations).
\item[3.] The metric of the spacetime satisfies the weak version of the twice-contracted second Bianchi identity everywhere, including in a neighborhood of any singularity of spacetime (of the type discussed in  item 2 above), thus allowing the usual conservation laws of total (i.e. field+particle) energy, momentum, and angular momentum to hold in a weak sense.
\end{itemize}
In Part I, we showed how these principles may be used to derive a classical equation of motion for world-lines of matter particles (which are identified with time-like singularities of spacetime) based entirely on momentum balance laws.  

Our approach to a quantum law of motion begins by replacing the third item above with another set of hypotheses\footnote{We emphasize that this is a preliminary list based on what we know so far, and  is subject to modification and expansion as our investigations continue.} :
\begin{itemize}
\item[3'.] The quantum-mechanical wave function of a single particle is a section of a Clifford-algebra-valued spin bundle over configuration spacetime. Different species of particles are distinguished by their wave functions at a point in spacetime belonging to different {\em generalized ideals} (stable subspaces) of that Clifford algebra\footnote{See Furey \cite{Fur2015,Fur2018} for details on this point of view.}.

\item[4'.] The wave function of a system of $N$ relativistic particles is defined on the $N$-body configuration spacetime, and takes its values in a tensor space generated by the tensor product of bases for the constituent 1-body wave function spaces.

\item[5'.] The wave function of a single particle satisfies a relativistic wave equation that is equivariant with respect to the action of the Poincar\'e group on the bundle of frames for the tangent space at any point in spacetime.

\item[6'.] As a consequence of satisfying that wave equation, the wave function of a single particle possesses a conserved, future-directed, timelike (or causal, if the particle is massless), and normalizable current, whose component in the normal direction to a given spacelike foliation of spacetime, when evaluated at a point on a leaf of that foliation gives the probability density of the particle trajectory crossing the leaf at that point (by the Born rule.) The same holds for the $N$-particle version of the wave function and an $N$-particle tensor current that is jointly conserved.
\item[7'] The motion of a system of $N$ particles is guided by a multi-time wave function defined on the $N$-particle configuration spacetime, in accordance with a relativistic generalization of the de Brogile-Bohm guiding law, such as the Hypersurface Bohm-Dirac Law, in which a conserved $N$-particle tensor current  evaluated at the actual positions of those $N$-particles determines their actual velocities.
\item[8'] In the classical limit, the quantum motions converge to those described in part I.
\end{itemize}

\section{Least Invasive Quantization}
  \label{sec:LIQ}
In this section we demonstrate by way of a very simple example, how it may be possible to continuously deform a classical law into a quantum law of motion for the same particle.  

Consider a particle of (bare) mass $m$ in an external potential $V = V(t,q)$, $q \in \mathbb{R}^d$.  The relativistic Lagrangian for the particle's motion is 
\beq \ell(t,q,\dot{q}) = m(1-\sqrt{1-\dot{q}^2}) - V(t,q). \eeq
The corresponding action is 
\beq \cA[q] := \int_{t_1}^{t_2} \ell(t,q,\dot{q}) dt.\eeq
We define the canonical momentum to be 
\beq 
p := \frac{\p \ell}{\p \dot{	q}} = \frac{m \dot{q}}{\sqrt{1-\dot{q}^2}},
\eeq
so the Hamiltonian is
\beq H(t,q,p) = p \dot{q} - \ell = \sqrt{m^2 + p^2} - m + V(t,q). \eeq
The Hamilton equations are therefore
\beq
 \dot{q} = \frac{\p H}{\p p} = \frac{p}{\sqrt{m^2+p^2}},\qquad
\dot{p} = - \frac{\p H}{\p q} = - \p_q V(t,q).
\eeq
%
The {\em Hamilton-Jacobi equation}
\beq 
\frac{\p S}{\p t} + H(q, \frac{\p S}{\p q},t) = 0,
\eeq
 is a first-order nonlinear PDE for the {\em Hamilton-Jacobi phase function}  $S = S(t,q)$. It arises in the context of solving Hamilton's equations by finding a {\em canonical transformation}, i.e. a change of variable in phase space $(q,p)$ such that the transformed Hamiltonian is constant.   We note that $S$ is defined on the space of generic positions of the particle, i.e. it's a function defined on the configuration space (or configuration spacetime, in the relativistic setting.)

In the case of our particular example, it's easy to see that the Hamilton-Jacobi equation has the form of an eikonal equation:
\beq\label{eq:HJeik}
|\p_t \tilde{S} + V|^2 - |\p_q \tilde{S}|^2  = m^2,\qquad \tilde{S} := S - mt.
\eeq
Moreover, this equation is derivable from an action principle:  Let $\varrho \geq 0$ be another function defined on the configuration spacetime,  set
\beq
\cL := \varrho \left( |\p_t \tilde{S} + V|^2 - |\p_q \tilde{S}|^2  - m^2\right),
\eeq
and let $\cS := \iint \cL dq dt$ be the corresponding action.  It is then obvious that the stationary points of $\cS$ with respect to compactly supported variations in $\varrho$ satisfy \refeq{eq:HJeik}.  If we now let $\varrho$  be an active variable, and set the variation of $\cS$ with respect to $\tilde{S}$ also equal to zero, we obtain the following continuity-type equation for $\varrho$:
\beq \label{eq:u}
\p^\mu (\varrho u_\mu) = 0.
\eeq
Here $\mu = 0,\dots,d$, indices are raised and lowered using the Minkowski metric $\eta = \mbox{diag}(1,-1,\dots,-1)$ on the configuration spacetime, and we are employing the Einstein summation convention.  The covectorfield $u$ is by definition
\beq
u_0 := \p_t \tilde{S} + V,\qquad u_i = \p_{q^i} \tilde{S},
\eeq
so that by \refeq{eq:HJeik}, 
\beq u_\mu u^\mu = m^2. \eeq

Let us now define a complex-valued field on the configuration spacetime
\beq \psi := \sqrt{\varrho} e^{i\tilde{S}/\hbar}. \eeq
Evidently, using that $(\varrho, \tilde{S})$ satisfy the Euler-Lagrange equations for the stationary points of the action $\cS$, one can derive the equation that $\psi$ must satisfy, which will be a nonlinear PDE.  

The goal of a {\em least invasive quantization} procedure is to find a continuous deformation of the Lagrangian $\cL$, now thought of as $\cL[\psi]$, to a {\em nearby} Lagrangian $\tilde{\cL}[\psi] = \cL[\psi] + \hbar^2\cL'[\psi]$ in such a way that the Euler-Lagrange equation of the new action corresponding to $\tilde{\cL}$ is a {\em linear} PDE in $\psi$.  This is easily accomplished by adding a {\em Fisher-like} term to $\cL$: Let
\beq
\tilde{\cL} := \cL + \hbar^2 \frac{\p_\mu \varrho \p^\mu \varrho}{4 \varrho}.
\eeq
A computation then shows that
\beq 
\tilde{\cL}[\psi] = \hbar^2|( \p_t +\frac{i}{\hbar}V)\psi|^2 - \hbar^2|\p_q \psi|^2 + m^2 |\psi|^2 
\eeq
so that a stationary point $\psi$ will satisfy the linear Klein-Gordon equation
\beq
\left(D^\mu D_\mu + m^2\right)\psi = 0,\qquad D_\mu := -i\hbar\p_\mu + A_\mu,
\eeq
which (under appropriate assumptions relating $A_\mu$ and $V$), is a Lorentz-covariant equation for a massive spin-0 particle.  

It should be noted that in  the above admittedly over-simplified setting, $V$ was taken to be an ``external" field, but in order to connect with our program of finding equations of motions for singularities of a dynamic field, $V$ needs to be determined by the field for which the particle in question is a singularity.   For example, if we are to identify the electron with certain singularities of the electromagnetic field, then the corresponding $V$ in the wave equation for the electron would have to be determined in terms of that electromagnetic field, evaluated at generic positions of the electron.  For a detailed discussion of a proper least invasive quantization in the electromagnetic context see \cite{Kie04b}. In any case, since the electron is a spin-1/2 particle, the appropriate wave equation is not Klein-Gordon's but Dirac's, so the above process needs to be suitably modified.  

Finally, we should note that even though in \refeq{eq:u} one may formally set $\rho := \varrho u^0$ and $v^k :=  u^k/u^0$ to obtain what looks like a continuity equation for a probability density
\beq \p_t \rho + \p_k (\rho v^k) = 0
\eeq
and proceed to derive a Born Rule and guiding law for the particle from it, there is a-priori no reason why $\rho$ would be non-negative.  This suggests that even for spin-0 particles, the Klein-Gordon may not be the correct equation. Indeed, the lack of a future-directed timelike conserved current for solutions of the Klein-Gordon equation is a main reason for researchers doubting whether bosons can be treated as particles.  We believe they can, and have already addressed this issue for photons \cite{KTZ2018}. The question of what the correct wave equation is for spin-0 bosons will be addressed in a forthcoming publication of ours.

\section{Quantum Law of Motion for Point Test Particles}
\subsection{Point particle in Hoffmann-like spacetimes}
We now move on to a  more realistic scenario, of a single point charge, thought of as the only time-like singularity of a static, spherically symmetric 4-dimensional electrovacuum spacetime, i.e. a solution of the Einstein-Maxwell equations.  As was shown in \cite{BKTZ2019}, the case where the electromagnetic vacuum law is the standard, linear law of Maxwell ($D=E, H=B$) yields a spacetime that is too singular for weak Bianchi identity to be satisfied at the singularity, so that our third working hypothesis is not satisfied.  We showed however, that there are {\em nonlinear} vacuum laws, including the one proposed by Born \& Infeld \cite{BI1933}, for which the corresponding static spherically symmetric solution of the Einstein-Maxwell equations (which was fully analyzed in \cite{TZ2011}) {\em does} satisfy the weak Bianchi identity. The spacetime that corresponds to the Born-Infeld vacuum law was discovered by Hoffmann \cite{Hof1935} in 1935. 

As a first step towards a full-blown two-body quantum-mechanical problem, one may study the dynamics of a single electron, thought of as a {\em test} point charge, placed in the electrostatic and gravitational field of another, much more massive point charge such as a nucleus, the latter thought of as the central singularity of a static spherically symmetric spacetime, while the contribution of the former to the geometry of the spacetime is ignored. At the same time, the Born-Oppenheimer approximation allows one to reduce the 2-body quantum mechanical situation to a single body one. The dynamics of the electron is thus determined by its wave function that, due to the electron being a massive spin-1/2 particle, satisfies the Dirac equation on the background spacetime of the nucleus, with the electrostatic field of the nucleus appearing in the Dirac operator as a minimal coupling term.  

This program was carried out successfully in \cite{Moulik2015}, where it was shown that the Dirac Hamiltonian on Hoffmann-like spacetimes is essentially self-adjoint regardless of the nuclear charge, its essential spectrum is the same as that of the free Dirac operator on Minkowski space, its point spectrum is non-empty and consists of infinitely many eigenvalues in the gap, accumulating at the right endpoint.  Numerical investigations of this problem are in the works, including comparison of the eigenvalues with those of the Dirac+Coulomb potential and the Dirac+Born electrostatic potential Hamiltonians on a Minkowski background.   
\subsection{Point particle in zero-gravity Kerr-Newmann spacetime}
\begin{figure}[h]
\centering
\input{zgkn3d.tex}
\caption{\label{fig:zgkn3d} Visualizing the double-sheeted spacelike slices of zero-$G$ Kerr-Newmann spacetime as a branched cover of the un-knot. Arrows indicate that the top of the disk in one copy of $\mathbb{R}^3$ is to be identified with the bottom of the disk in the other copy.}
\end{figure}
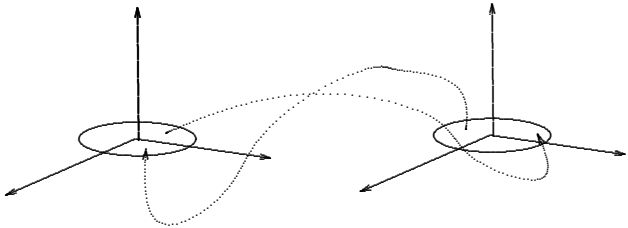

As mentioned above, the main obstacle on the path to a well-defined quantum theory that could encompass both gravity and electromagnetism is that when Maxwell-Lorentz electrodynamics is coupled to Einstein's gravity, the infinities inherent in ML cause the spacetime to have curvature singularities that are too strong for even a weak notion of energy-momentum conservation to hold.   In our quest to find a remedy for the strong curvature singularities of well-known solutions of Einstein-Maxwell equations such as the Kerr-Newmann solution, another avenue that we have pursued is to work with the {\em zero-gravity limit} of such singular spacetimes.  This is a (geometrically well-defined) limit when Newton's gravitational constant $G$, which appears as a parameter in these metrics, is set to zero.  The limiting spacetime and the electromagnetic field on it were analyzed in \cite{TZ2014}.  It is axially symmetric, static, and locally isometric to Minkowksi space, but is topologically non-trivial. These facts were already known to Carter \cite{Car68}, who discovered the maximal analytical extension of the Kerr-Newmann solution.   It was also known that the constant time slices of this manifold are double-sheeted, and have the topology of $\mathbb{R}^3$ branched over the un-knot (See Fig.~\ref{fig:zgkn3d}.) 

The spacetime is indeed singular on a 2-dim. timelike cylinder $\mathbb{S}^1\times \mathbb{R}$, which is the world-tube of a ring-like singularity. The most intriguing aspect of this solution is that at any instant of time the ring singularity, when viewed from one of the two sheets of space, appears to be positively charged, and from the other sheet, negatively charged.  

\begin{figure}[h]
\centering
\includegraphics[width=100pt,height=100pt]{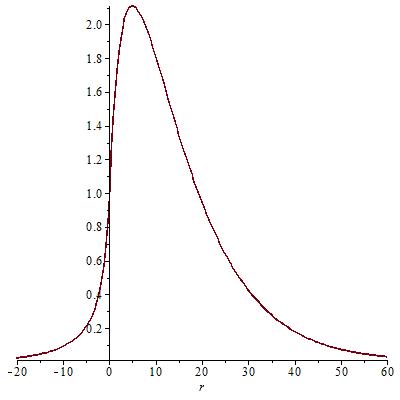}
\caption{\label{fig:tigerstail} Ground state of single-electron Dirac Hamiltonian on a zGKN background}
\end{figure}

As a first step in studying the quantum problem, one may identify the zero-gravity Kerr-Newmann (zGKN) solution with the spacetime outside a ring-like  singularity representing a positively charged particle such as a proton.  Since the metric of zGKN and the electromagnetic field corresponding to it are well known, it is easy to formulate the quantum dynamics of a test electron placed in the vicinity of the ring singularity by studying the Dirac equation on the zGKN background.  This task was carried out in \cite{KTZ2015}, where it was shown, using techniques developed in \cite{BSW2005,WinYam2006}, that the pertinent Dirac Hamiltonian is essentially self-adjoint and its essential spectrum is the same as the standard Dirac operator on Minkowski space.  We further showed that its discrete spectrum is nonempty, provided the radius of the ring is small enough. The ground state of the Dirac Hamiltonian for the single test electron has support in both sheets of the spacetime, but is mainly concentrated in the sheet where the ring singularity appears to be positively charged, see Fig.~\ref{fig:tigerstail}, where the horizontal axis is a radial coordinate that runs from $-\infty$ to $0$ in one sheet and $0$ to $\infty$ in the other sheet, while the vertical axis is $|\psi|^2$. 

\begin{figure}[h]
\centering
\includegraphics[width=150pt,height=100pt]{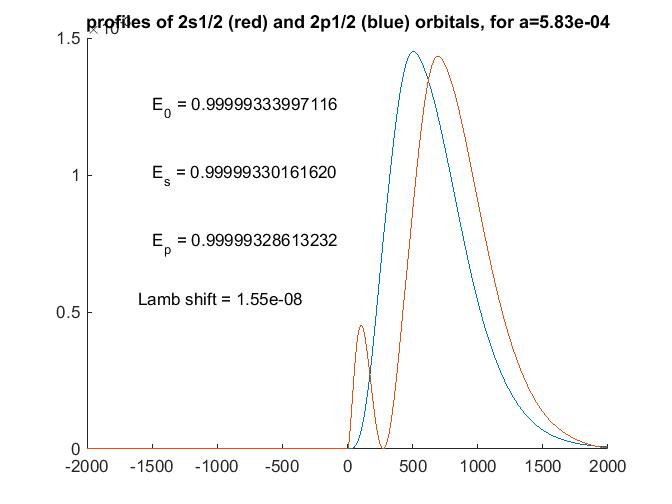}

\caption{\label{fig:lambshift} Numerical evidence of a qualitative Lamb shift for Dirac's Hamiltonian on zGKN.  The prameter $a$ is the ring radius, which here is set equal to electron's anomalous magnetic moment, measured in units of the Compton wavelength of electron.}
\end{figure}

Further analysis showed that  the radius of the ring should be of the order of the {\em anomalous magnetic moment of the electron}, and that given this choice, the full spectrum will be close to the standard Dirac+Coulomb problem for Hydrogenic atoms.  Furthermore, the broken symmetry due to nonzero ring radius causes the well-known degeneracy in the Dirac spectrum of Hydrogen to be broken, resulting in effects that are qualitatively similar to Lamb shift and hyperfine splitting, without the need to appeal to QED methods, see Fig.~\ref{fig:lambshift} (axes have the same meaning as in Fig.~\ref{fig:tigerstail}.)

\section{Towards a Quantum Law of Motion for Ring-Like Singularities of Spacetime}
\begin{figure}[h]
\centering
\includegraphics[scale=0.5]{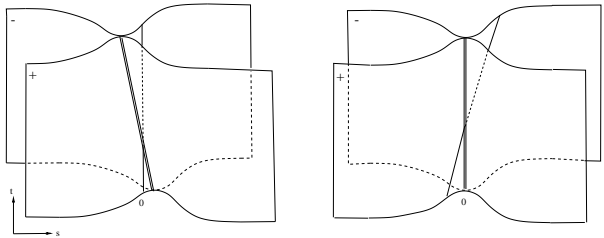}
\caption{\label{fig:twoviews} Left: The  physical spacetime with a single ring singularity in straight line motion relative to the rest frame of a designated origin, which in the depicted scenario gets ``swept over'' by the ring singularity;
Right: Trajectory of a test particle in straight line motion in the static z$G$KN spacetime having the center of its ring singularity 
as the origin (marked $0$). The test particle transits through the ring from one sheet to the other.}
\end{figure}

Going back to the intriguing feature of the ring singularity of zGKN, i.e. that the same ring is found to have a positive charge when seen from one of the two sheets of that spacetime, and a negative charge when seen from the other sheet, one may wonder whether this ring singularity is representing both a particle and its anti-particle.  In \cite{KTZ2016}, extrapolating from the existence of such ``bi-particle" structures in General Relativity, we proposed a
novel interpretation of Dirac's ``wave equation for the relativistic
electron" in which
the electron and the positron are merely the two different ``topological spin"
states of a single more fundamental particle, not distinct particles in their own
right. This novel
interpretation resolves the dilemma that Dirac's wave equation seems to be
capable of describing both the electron and the positron in ‘external’ fields in
many relevant situations, while the bi-spinorial wave function has only a
single position variable in its argument, not two --as it should if it were a
quantum-mechanical two-particle wave equation.  We formulated a Dirac equation for such a ring-like bi-particle which interacts with a static point
charge located elsewhere in the topologically non-trivial physical space
associated with the moving ring particle.  Indeed, for quasi-static motions of the ring, this equation is nothing but the one we have been discussing in the previous section, namely Dirac's equation on the zGKN background!  The difference is in the interpretation, since the wave function in the present case is defined on the configuration space of the center of the ring, which is now moving with respect to a fixed point charge located on one of the two sheets.  See Fig.~\ref{fig:twoviews} for a schematic illustration of the two interpretations. Furthermore, we showed that
 the motion of the ring can be governed by a de Broglie-Bohm type law extracted from the Dirac equation. 

Our work in this direction is so far limited to quasi-static motions of the ring, which result in the ring remaining circular to the leading order, so that its motion can be specified solely in terms of the trajectory of the center of the circle, together with the evolving unit normal to the plane of the circle.  Clearly, much work remains to be done in order to extend these results to the general case, which would presumably involve deformations of the ring as it moves.

The next natural step in this direction would be to study the full 2-body quantum problem, either in the context of nonlinear electrodynamics, or in the zero-gravity context by studying the two-rings analog of zGKN.  Due to the strong nonlinearity of the BI vacuum law however, there is currently no explicit solution known to the {\em classical} 2-body problem, even in the absence of gravity, thus making any progress in that direction quite difficult.    As to the two rings, that analysis is already under way.  We know for example that the space should be four-sheeted (in general, $2^N$ sheets are needed for $N$ ring singularities), and that it is possible to come up with a Dirac equation for the 2-body wave function (with synchronized times), equipped with a minimal coupling term that models the electromagnetic interaction of the two rings.  It is important to note however, that there is no chance of ``turning gravity back on" in these models, since doing so will bring back the strong singularities of the type present in the Kerr-Newman spacetime, together with all the causal pathologies that they entail.

As described in Part I of this report, there are other electromagnetic theories, such as the Bopp-Lande-Thomas-Podolsky (BLTP) theory, which may allow us to find electromagnetic spacetimes whose singularities are mild enough so that they can represent charged particles.   Once such solutions are known, a least invasive quantization procedure as outlined in Section~\ref{sec:LIQ} could conceivably lead to a quantum law of motion for those particles.  Investigations are currently under way on various aspects of this program.

\section{Quantum Laws of Motion for Interacting Point-Particle Systems}

As mentioned at the outset, our goal is the eventual formulation of a relativistic quantum-mechanical theory of motion for a fixed number $N$ of electrons, photons, and their anti-particles in 3+1-dimensional Minkowski 
spacetime, and its generalization to other particles and interactions.  We are certainly not the first researchers to state this as our goal, as can be attested to by the following passage in \cite{Dar1932}:
\begin{quote}\small
``The Compton effect, at its discovery, was regarded as a simple collision of two bodies, and yet the detailed discussion at the present time involves the idea of the annihilation of one photon and the simultaneous creation of one among an infinity of other possible ones. We would like to be able to treat the effect as a two-body problem, with the scattered photon regarded as the same individual as the incident, in just the way we treat  the collisions of electrons.''
\end{quote}
Such a goal however, of treating electrons and photons on an equal footing,
within a quantum-{\em mechanical} framework of a fixed number of particles,  has so far remained elusive.
 Progress has been obstructed in particular by the lack of a viable candidate for the quantum-mechanical photon wave function and 
its pertinent relativistic wave equation which furnishes a conserved probability current for the photon position, obeying Born's rule.
 Recently such a photon wave function and wave equation have been constructed in \cite{KTZ2018}. 
 In a forthcoming joint work with M. Lienert \cite{KLTZ2019}, we show how the photon wave equation of \cite{KTZ2018} can be coupled with Dirac's well-known relativistic wave equation 
for the electron in a Lorentz-covariant manner to accomplish what Darwin has asked us to do: to ``treat the effect as a two-body 
problem, with the scattered photon regarded as the same individual as the incident, in just the way we treat the collisions of electrons.'' 

 We work with Dirac's manifestly Lorentz-covariant formalism of multi-time wave functions \cite{Dir1932}.
 For our $N=2$ body problem the wave function $\Psi(\bx_\ph,\bx_\el)$ depends on the two generic
spacetime events $\bx_\el$ and $\bx_\ph$ of the electron and the photon, respectively, which must be space-like separated. 
 Both the Dirac operator of a free electron \cite{ThallerBOOK}, and the Dirac-type operator of a free photon constructed in \cite{KTZ2018},
act on $\Psi(\bx_\ph,\bx_\el)$.
 Unique solvability of this system of evolution equations requires imposing a suitable {\em boundary condition} at the subset of
co-incident events, $\{\bx_\el=\bx_\ph\}$. 
 Conservation of the particle currents dictates the boundary condition, up to a choice in a phase. 
 We prove that the resulting initial-boundary-value problem is well-posed.

\begin{figure}[ht]
\centering
\includegraphics[width=60pt,height=60pt]{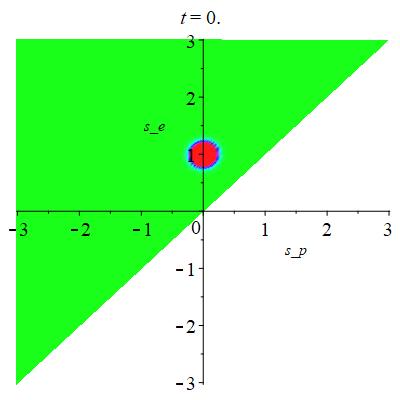}
\includegraphics[width=60pt,height=60pt]{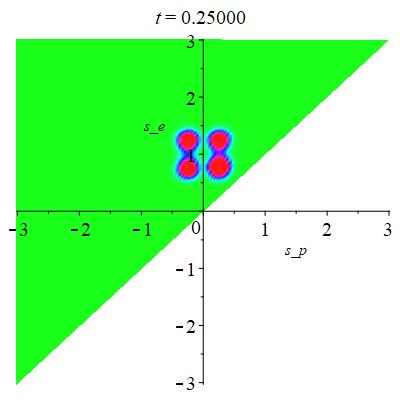}
\includegraphics[width=60pt,height=60pt]{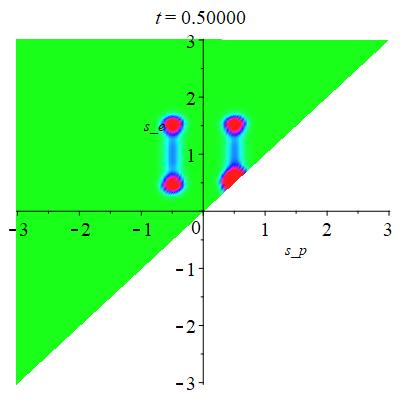}
\includegraphics[width=60pt,height=60pt]{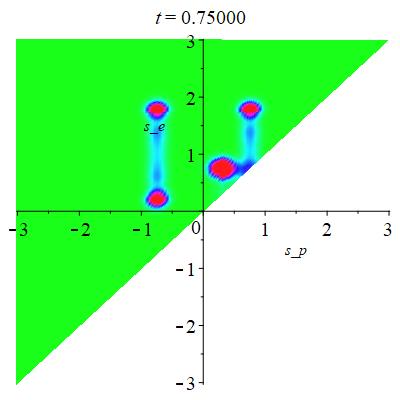}
\includegraphics[width=60pt,height=60pt]{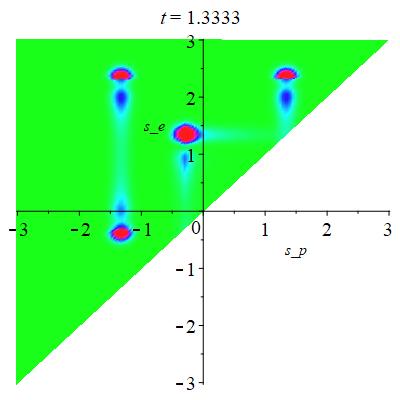}
\caption{\label{fig:movie} Contour plot of the density $\rho$ in photon-electron configuration space
at six consecutive snapshots of common time $t=t_\el=t_\ph$. The photon axis is horizontal, the electron axis vertical.
}
\end{figure}

 It is intuitively obvious that a boundary condition at co-incident events $\{\bx_\el=\bx_\ph\}$ 
amounts to a local pair interaction between electron and photon in a Lorentz-covariant manner. 
 Our boundary condition is compatible with the kind of interaction expected for an electron and a photon in Compton scattering.  In particular, when we plot the joint 2-body probability density $\rho$ as a function of generic positions of the two particles, we observe that, as it evolves in time, the density forms four peaks, corresponding to the wave function being mostly supported near four distinct configurations: (1) both particles going to the left, (2) both particles going to the right, (3) particles going away from each other for all time; and (4) particles initially moving towards each other and ``bouncing off" of one another.   Indeed the fourth peek appears to ``hit" the boundary and get reflected, see Fig.~\ref{fig:movie} for snapshots of the evolving density\footnote{You can see an animation of the density plot of $\rho$ as a function of the two positions by going to \url{http://sites.math.rutgers.edu/~shadi/phel1dint_slow.gif}
}. 

 We have also carried out a limited number of numerical experiments with our system of equations, which indeed demonstrate the process of Compton scattering, 
but which also have revealed an unexpected novel phenomenon: photon capture and subsequent release by the electron.
 For all practical purposes this scenario seems indistinguishable from the scenario of an annihilation of a photon, followed by a subsequent
emission of another one. 
 In our quantum-mechanical $N=2$-body model the photon of course never gets destroyed or created, precisely as envisioned by Darwin. 

 Beyond demonstrating that relativistic quantum mechanics with a fixed finite number $N$ of interacting particles is feasible, 
we also formulate the de Broglie--Bohm-type foundations of this Lorentz-covariant quantum model\footnote{This should put to rest
the often-voiced claim that the de Broglie--Bohm theory could not be made relativistic.}.
 This is accomplished by adapting to our interacting 2-body model the so-called ``hypersurface {Bohm--Dirac}''-type formulation for non-interacting particles 
\cite{DGMZ1999}.  This formulation requires one to  specify a foliation of spacetime by spacelike hypersurfaces.   In our model these hypersurfaces are given by a time-like Killing vector field   that is determined self-consistently 
by the initial data of the wave function.
 The guiding law for the particles is furnished by the conserved current of our quantum-mechanical multi-time wave function. 
 We extend a theorem of Teufel--Tumulka \cite{TT2005}, which implies that unique particle motions typically exist globally in time.

\begin{figure}[h]
\centering
\includegraphics[width=100pt,height=100pt]{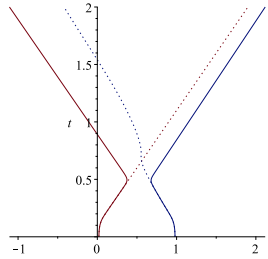}
\caption{\label{fig:intvsnonint} Iinteracting (solid) versus non-interacting (dotted) electron-photon trajectories.}
\end{figure}

 The numerically computed trajectories confirm that our boundary condition causes the particles to scatter off of one another.  See Fig.~\ref{fig:intvsnonint} for example of a two pairs of particles with identical initial conditions, one evolved according to the free evolution (i.e. without any boundary condition on the coincidence set) and the other using our boundary condition.

\begin{figure}[h]
\centering
\includegraphics[width=100pt,height=100pt]{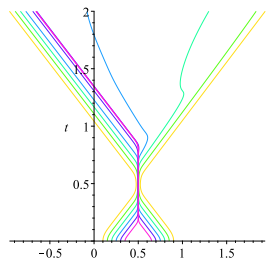}
\caption{\label{fig:romance}World-lines corresponding to possible capture-and-release phenomena (Different colors correspond to the photon-electron pair starting at different initial positions that are successively closer to each other.)}
\end{figure}

 The intriguing phenomenon of photon ``capture and release'' by the electron that we alluded to in the above can also be illustrated by the numerically 
computed electron and photon trajectories, see Fig.~\ref{fig:romance}.

\begin{figure}[h]
\centering
\includegraphics[width=100pt,height=100pt]{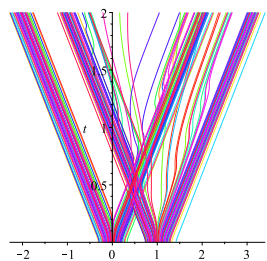}
\caption{\label{fig:bohmtraj}Electron \&\ photon world-lines corresponding to 
100 pairs of
 typical initial positions.}
\end{figure}

 Finally, by sampling a large ensemble of random positions distributed according to the initial wave function,  we can also illustrate that the empirical statistics over the possible actual 
trajectories reproduces Born's rule in our model, a consequence of the equivariance of the evolution of the probability densities.  See Fig.~\ref{fig:bohmtraj} and note the four clusters of trajectories that correspond to the four peaks of the joint probability density that was mentioned in the above.

More numerical investigation is currently under way, to explore various regions in the parameter space for the dynamics (e.g. mass and energy of incident electron, frequency profile of incident photon, their initial distance, etc.) and to quantify the energy and momentum transfer between the two particles.

\section*{Acknowledgements} We gratefully acknowledge helpful discussions with:
M. K. Balasubramanian,
E. Carlen,
O. Darrigol,
C. Furey,
S. Goldstein,
J.-M. Graf,
F. Hehl,
M. Jansen,
H. Jauslin,
J. Lebowitz,
M. Lein,
N. Leopold,
M. Lienert,
T. Newman,
B. Simon,
A. Soffer,
H. Spohn,
W. Struyve,
R. Tumulka,
and M. Winklmeier.
 We thank the organizers, R. Ruffini and R. Jantzen, for inviting us to MG15.

\bibliographystyle{plain}

\end{document}

%% file: zgkn3d.tex
%
%
\font\thinlinefont=cmr5
\begingroup\makeatletter\ifx\SetFigFont\undefined%
\gdef\SetFigFont#1#2#3#4#5{%
  \reset@font\fontsize{#1}{#2pt}%
  \fontfamily{#3}\fontseries{#4}\fontshape{#5}%
  \selectfont}%
\fi\endgroup%
\mbox{\beginpicture
\setcoordinatesystem units <0.50000cm,0.50000cm>
\unitlength=0.50000cm
\linethickness=1pt
\setplotsymbol ({\makebox(0,0)[l]{\tencirc\symbol{'160}}})
\setshadesymbol ({\thinlinefont .})
\setlinear
%
%
\linethickness= 0.500pt
\setplotsymbol ({\thinlinefont .})
{\color[rgb]{0,0,0}\ellipticalarc axes ratio  1.558:0.447  360 degrees 
	from  6.066 21.177 center at  4.508 21.177
}%
%
%
\linethickness= 0.500pt
\setplotsymbol ({\thinlinefont .})
{\color[rgb]{0,0,0}\plot  4.540 21.145  4.508 24.670 /
%
%
\plot  4.574 24.416  4.508 24.670  4.447 24.415 /
}%
%
%
\linethickness= 0.500pt
\setplotsymbol ({\thinlinefont .})
{\color[rgb]{0,0,0}\plot  4.445 21.177  1.016 19.685 /
%
%
\plot  1.224 19.845  1.016 19.685  1.274 19.728 /
}%
%
%
\linethickness= 0.500pt
\setplotsymbol ({\thinlinefont .})
{\color[rgb]{0,0,0}\plot  4.477 21.177  8.033 20.669 /
%
%
\plot  7.772 20.642  8.033 20.669  7.790 20.768 /
}%
%
%
\linethickness= 0.500pt
\setplotsymbol ({\thinlinefont .})
{\color[rgb]{0,0,0}\ellipticalarc axes ratio  1.558:0.447  360 degrees 
	from 15.496 21.273 center at 13.938 21.273
}%
%
%
\linethickness= 0.500pt
\setplotsymbol ({\thinlinefont .})
{\color[rgb]{0,0,0}\plot 13.970 21.241 13.938 24.765 /
%
%
\plot 14.004 24.512 13.938 24.765 13.877 24.510 /
}%
%
%
\linethickness= 0.500pt
\setplotsymbol ({\thinlinefont .})
{\color[rgb]{0,0,0}\plot 13.875 21.273 10.446 19.780 /
%
%
\plot 10.653 19.940 10.446 19.780 10.704 19.823 /
}%
%
%
\linethickness= 0.500pt
\setplotsymbol ({\thinlinefont .})
{\color[rgb]{0,0,0}\plot 13.906 21.273 17.462 20.765 /
%
%
\plot 17.202 20.738 17.462 20.765 17.220 20.863 /
}%
%
%
\linethickness= 0.500pt
\setplotsymbol ({\thinlinefont .})
{\color[rgb]{0,0,0}\plot  4.540 21.145  4.508 24.670 /
%
%
\plot  4.574 24.416  4.508 24.670  4.447 24.415 /
}%
%
%
\linethickness= 0.500pt
\setplotsymbol ({\thinlinefont .})
\setdots < 0.1270cm>
{\color[rgb]{0,0,0}\plot 13.240 21.431 13.240 21.433 /
\plot 13.240 21.433 13.242 21.440 /
\plot 13.242 21.440 13.242 21.450 /
\plot 13.242 21.450 13.244 21.467 /
\plot 13.244 21.467 13.246 21.488 /
\plot 13.246 21.488 13.250 21.518 /
\plot 13.250 21.518 13.255 21.556 /
\plot 13.255 21.556 13.257 21.598 /
\plot 13.257 21.598 13.261 21.647 /
\plot 13.261 21.647 13.263 21.702 /
\plot 13.263 21.702 13.265 21.761 /
\plot 13.265 21.761 13.265 21.825 /
\plot 13.265 21.825 13.263 21.891 /
\plot 13.263 21.891 13.259 21.958 /
\plot 13.259 21.958 13.250 22.026 /
\plot 13.250 22.026 13.240 22.096 /
\plot 13.240 22.096 13.225 22.164 /
\plot 13.225 22.164 13.206 22.231 /
\plot 13.206 22.231 13.180 22.299 /
\plot 13.180 22.299 13.151 22.365 /
\plot 13.151 22.365 13.113 22.430 /
\plot 13.113 22.430 13.066 22.494 /
\plot 13.066 22.494 13.011 22.555 /
\plot 13.011 22.555 12.948 22.614 /
\plot 12.948 22.614 12.871 22.672 /
\plot 12.871 22.672 12.783 22.727 /
\plot 12.783 22.727 12.683 22.775 /
\plot 12.683 22.775 12.571 22.822 /
\plot 12.571 22.822 12.446 22.860 /
\plot 12.446 22.860 12.347 22.883 /
\plot 12.347 22.883 12.243 22.904 /
\plot 12.243 22.904 12.139 22.921 /
\plot 12.139 22.921 12.040 22.936 /
\plot 12.040 22.936 11.940 22.949 /
\plot 11.940 22.949 11.847 22.959 /
\plot 11.847 22.959 11.760 22.970 /
\plot 11.760 22.970 11.680 22.979 /
\plot 11.680 22.979 11.606 22.987 /
\plot 11.606 22.987 11.538 22.995 /
\plot 11.538 22.995 11.477 23.004 /
\plot 11.477 23.004 11.422 23.012 /
\plot 11.422 23.012 11.373 23.021 /
\plot 11.373 23.021 11.328 23.031 /
\plot 11.328 23.031 11.288 23.040 /
\plot 11.288 23.040 11.252 23.048 /
\plot 11.252 23.048 11.218 23.057 /
\plot 11.218 23.057 11.189 23.067 /
\plot 11.189 23.067 11.157 23.074 /
\plot 11.157 23.074 11.127 23.082 /
\plot 11.127 23.082 11.098 23.089 /
\plot 11.098 23.089 11.066 23.095 /
\plot 11.066 23.095 11.032 23.099 /
\plot 11.032 23.099 10.996 23.101 /
\plot 10.996 23.101 10.954 23.099 /
\plot 10.954 23.099 10.907 23.097 /
\plot 10.907 23.097 10.854 23.091 /
\plot 10.854 23.091 10.795 23.080 /
\plot 10.795 23.080 10.725 23.067 /
\plot 10.725 23.067 10.649 23.048 /
\plot 10.649 23.048 10.562 23.025 /
\plot 10.562 23.025 10.465 22.993 /
\plot 10.465 22.993 10.359 22.957 /
\plot 10.359 22.957 10.240 22.913 /
\plot 10.240 22.913 10.111 22.862 /
\plot 10.111 22.862  9.972 22.803 /
\plot  9.972 22.803  9.823 22.735 /
\plot  9.823 22.735  9.667 22.657 /
\plot  9.667 22.657  9.502 22.572 /
\plot  9.502 22.572  9.335 22.479 /
\plot  9.335 22.479  9.193 22.394 /
\plot  9.193 22.394  9.053 22.308 /
\plot  9.053 22.308  8.915 22.217 /
\plot  8.915 22.217  8.782 22.123 /
\plot  8.782 22.123  8.653 22.030 /
\plot  8.653 22.030  8.530 21.935 /
\plot  8.530 21.935  8.414 21.842 /
\plot  8.414 21.842  8.306 21.751 /
\plot  8.306 21.751  8.202 21.660 /
\plot  8.202 21.660  8.107 21.571 /
\plot  8.107 21.571  8.018 21.482 /
\plot  8.018 21.482  7.938 21.397 /
\plot  7.938 21.397  7.861 21.315 /
\plot  7.861 21.315  7.791 21.234 /
\plot  7.791 21.234  7.728 21.156 /
\plot  7.728 21.156  7.671 21.080 /
\plot  7.671 21.080  7.618 21.006 /
\plot  7.618 21.006  7.567 20.934 /
\plot  7.567 20.934  7.523 20.862 /
\plot  7.523 20.862  7.480 20.792 /
\plot  7.480 20.792  7.440 20.724 /
\plot  7.440 20.724  7.402 20.657 /
\plot  7.402 20.657  7.364 20.591 /
\plot  7.364 20.591  7.328 20.523 /
\plot  7.328 20.523  7.294 20.458 /
\plot  7.294 20.458  7.258 20.392 /
\plot  7.258 20.392  7.220 20.326 /
\plot  7.220 20.326  7.182 20.259 /
\plot  7.182 20.259  7.144 20.193 /
\plot  7.144 20.193  7.101 20.125 /
\plot  7.101 20.125  7.055 20.058 /
\plot  7.055 20.058  7.008 19.988 /
\plot  7.008 19.988  6.955 19.920 /
\plot  6.955 19.920  6.900 19.848 /
\plot  6.900 19.848  6.839 19.778 /
\plot  6.839 19.778  6.773 19.706 /
\plot  6.773 19.706  6.703 19.634 /
\plot  6.703 19.634  6.629 19.560 /
\plot  6.629 19.560  6.549 19.488 /
\plot  6.549 19.488  6.464 19.416 /
\plot  6.464 19.416  6.375 19.346 /
\plot  6.375 19.346  6.282 19.279 /
\plot  6.282 19.279  6.185 19.211 /
\plot  6.185 19.211  6.085 19.149 /
\plot  6.085 19.149  5.984 19.090 /
\plot  5.984 19.090  5.882 19.037 /
\plot  5.882 19.037  5.781 18.993 /
\plot  5.781 18.993  5.683 18.955 /
\plot  5.683 18.955  5.560 18.919 /
\plot  5.560 18.919  5.446 18.900 /
\plot  5.446 18.900  5.342 18.893 /
\plot  5.342 18.893  5.249 18.904 /
\plot  5.249 18.904  5.165 18.925 /
\plot  5.165 18.925  5.091 18.957 /
\plot  5.091 18.957  5.027 18.999 /
\plot  5.027 18.999  4.970 19.050 /
\plot  4.970 19.050  4.921 19.111 /
\plot  4.921 19.111  4.879 19.177 /
\plot  4.879 19.177  4.843 19.251 /
\plot  4.843 19.251  4.811 19.332 /
\plot  4.811 19.332  4.786 19.416 /
\plot  4.786 19.416  4.765 19.505 /
\plot  4.765 19.505  4.748 19.598 /
\plot  4.748 19.598  4.735 19.696 /
\plot  4.735 19.696  4.724 19.795 /
\plot  4.724 19.795  4.716 19.895 /
\plot  4.716 19.895  4.710 19.996 /
\plot  4.710 19.996  4.707 20.098 /
\plot  4.707 20.098  4.705 20.195 /
\plot  4.705 20.195  4.705 20.292 /
\plot  4.705 20.292  4.707 20.384 /
\plot  4.707 20.384  4.710 20.470 /
\plot  4.710 20.470  4.712 20.549 /
\plot  4.712 20.549  4.714 20.621 /
\plot  4.714 20.621  4.718 20.686 /
\plot  4.718 20.686  4.720 20.739 /
\plot  4.720 20.739  4.724 20.786 /
\plot  4.724 20.786  4.727 20.822 /
\plot  4.727 20.822  4.729 20.849 /
\plot  4.729 20.849  4.731 20.892 /
\setsolid
%
%
\plot  4.781 20.635  4.731 20.892  4.655 20.641 /
\setdots < 0.1270cm>
}%
%
%
\linethickness= 0.500pt
\setplotsymbol ({\thinlinefont .})
{\color[rgb]{0,0,0}\plot  5.271 21.336  5.273 21.336 /
\plot  5.273 21.336  5.275 21.338 /
\plot  5.275 21.338  5.283 21.342 /
\plot  5.283 21.342  5.294 21.347 /
\plot  5.294 21.347  5.309 21.353 /
\plot  5.309 21.353  5.330 21.361 /
\plot  5.330 21.361  5.355 21.374 /
\plot  5.355 21.374  5.389 21.389 /
\plot  5.389 21.389  5.429 21.406 /
\plot  5.429 21.406  5.476 21.427 /
\plot  5.476 21.427  5.529 21.450 /
\plot  5.529 21.450  5.590 21.476 /
\plot  5.590 21.476  5.658 21.503 /
\plot  5.658 21.503  5.732 21.535 /
\plot  5.732 21.535  5.812 21.567 /
\plot  5.812 21.567  5.899 21.603 /
\plot  5.899 21.603  5.990 21.639 /
\plot  5.990 21.639  6.088 21.677 /
\plot  6.088 21.677  6.191 21.717 /
\plot  6.191 21.717  6.297 21.757 /
\plot  6.297 21.757  6.407 21.797 /
\plot  6.407 21.797  6.521 21.840 /
\plot  6.521 21.840  6.640 21.880 /
\plot  6.640 21.880  6.761 21.920 /
\plot  6.761 21.920  6.883 21.960 /
\plot  6.883 21.960  7.010 22.001 /
\plot  7.010 22.001  7.140 22.039 /
\plot  7.140 22.039  7.271 22.077 /
\plot  7.271 22.077  7.404 22.113 /
\plot  7.404 22.113  7.540 22.147 /
\plot  7.540 22.147  7.679 22.181 /
\plot  7.679 22.181  7.821 22.210 /
\plot  7.821 22.210  7.965 22.240 /
\plot  7.965 22.240  8.113 22.267 /
\plot  8.113 22.267  8.266 22.293 /
\plot  8.266 22.293  8.422 22.314 /
\plot  8.422 22.314  8.581 22.333 /
\plot  8.581 22.333  8.746 22.350 /
\plot  8.746 22.350  8.915 22.363 /
\plot  8.915 22.363  9.089 22.373 /
\plot  9.089 22.373  9.267 22.380 /
\plot  9.267 22.380  9.449 22.382 /
\plot  9.449 22.382  9.637 22.380 /
\plot  9.637 22.380  9.828 22.371 /
\plot  9.828 22.371 10.022 22.360 /
\plot 10.022 22.360 10.217 22.344 /
\plot 10.217 22.344 10.414 22.320 /
\plot 10.414 22.320 10.617 22.291 /
\plot 10.617 22.291 10.816 22.257 /
\plot 10.816 22.257 11.009 22.219 /
\plot 11.009 22.219 11.193 22.176 /
\plot 11.193 22.176 11.369 22.130 /
\plot 11.369 22.130 11.532 22.083 /
\plot 11.532 22.083 11.686 22.032 /
\plot 11.686 22.032 11.828 21.982 /
\plot 11.828 21.982 11.961 21.931 /
\plot 11.961 21.931 12.080 21.878 /
\plot 12.080 21.878 12.190 21.825 /
\plot 12.190 21.825 12.291 21.772 /
\plot 12.291 21.772 12.380 21.721 /
\plot 12.380 21.721 12.463 21.668 /
\plot 12.463 21.668 12.535 21.615 /
\plot 12.535 21.615 12.601 21.565 /
\plot 12.601 21.565 12.662 21.512 /
\plot 12.662 21.512 12.715 21.461 /
\plot 12.715 21.461 12.764 21.410 /
\plot 12.764 21.410 12.810 21.359 /
\plot 12.810 21.359 12.852 21.308 /
\plot 12.852 21.308 12.893 21.260 /
\plot 12.893 21.260 12.931 21.209 /
\plot 12.931 21.209 12.969 21.158 /
\plot 12.969 21.158 13.007 21.110 /
\plot 13.007 21.110 13.045 21.059 /
\plot 13.045 21.059 13.085 21.008 /
\plot 13.085 21.008 13.128 20.957 /
\plot 13.128 20.957 13.174 20.906 /
\plot 13.174 20.906 13.223 20.856 /
\plot 13.223 20.856 13.278 20.805 /
\plot 13.278 20.805 13.335 20.752 /
\plot 13.335 20.752 13.399 20.701 /
\plot 13.399 20.701 13.468 20.650 /
\plot 13.468 20.650 13.542 20.597 /
\plot 13.542 20.597 13.625 20.546 /
\plot 13.625 20.546 13.712 20.496 /
\plot 13.712 20.496 13.805 20.445 /
\plot 13.805 20.445 13.904 20.396 /
\plot 13.904 20.396 14.008 20.348 /
\plot 14.008 20.348 14.116 20.303 /
\plot 14.116 20.303 14.226 20.261 /
\plot 14.226 20.261 14.340 20.221 /
\plot 14.340 20.221 14.453 20.185 /
\plot 14.453 20.185 14.563 20.155 /
\plot 14.563 20.155 14.669 20.130 /
\plot 14.669 20.130 14.787 20.108 /
\plot 14.787 20.108 14.895 20.094 /
\plot 14.895 20.094 14.992 20.089 /
\plot 14.992 20.089 15.077 20.091 /
\plot 15.077 20.091 15.149 20.102 /
\plot 15.149 20.102 15.212 20.119 /
\plot 15.212 20.119 15.265 20.140 /
\plot 15.265 20.140 15.310 20.168 /
\plot 15.310 20.168 15.344 20.199 /
\plot 15.344 20.199 15.371 20.235 /
\plot 15.371 20.235 15.392 20.276 /
\plot 15.392 20.276 15.407 20.320 /
\plot 15.407 20.320 15.416 20.367 /
\plot 15.416 20.367 15.420 20.417 /
\plot 15.420 20.417 15.420 20.468 /
\plot 15.420 20.468 15.414 20.523 /
\plot 15.414 20.523 15.405 20.580 /
\plot 15.405 20.580 15.395 20.637 /
\plot 15.395 20.637 15.380 20.695 /
\plot 15.380 20.695 15.363 20.754 /
\plot 15.363 20.754 15.346 20.811 /
\plot 15.346 20.811 15.327 20.868 /
\plot 15.327 20.868 15.306 20.925 /
\plot 15.306 20.925 15.287 20.978 /
\plot 15.287 20.978 15.265 21.029 /
\plot 15.265 21.029 15.246 21.078 /
\plot 15.246 21.078 15.227 21.120 /
\plot 15.227 21.120 15.210 21.160 /
\plot 15.210 21.160 15.196 21.194 /
\plot 15.196 21.194 15.183 21.224 /
\plot 15.183 21.224 15.170 21.249 /
\plot 15.170 21.249 15.162 21.268 /
\plot 15.162 21.268 15.155 21.283 /
\plot 15.155 21.283 15.145 21.304 /
\setsolid
%
%
\plot 15.315 21.105 15.145 21.304 15.202 21.049 /
\setdots < 0.1270cm>
}%
\linethickness=0pt
\putrectangle corners at  0.991 24.790 and 17.488 18.868
\endpicture}